\documentclass[aps,superscriptaddress,showpacs,floatfix]{revtex4}
\usepackage{amsmath,amssymb,amsfonts,bbm,graphicx,hyperref,color,times}
\usepackage{subfigure}
\usepackage[english]{babel}
\newcommand{\bra}[1]{\ensuremath{\langle #1|}}
\newcommand{\ket}[1]{\ensuremath{|#1 \rangle}}
\newcommand{\braket}[2]{\ensuremath{\langle #1| #2 \rangle}}
\newcommand{\ketbra}[2]{\ensuremath{| #1 \rangle\hspace{-2pt} \langle #2 |}}
\newcommand{\braoket}[3]{\ensuremath{\langle #1 | #2 | #3 \rangle}}

\newcommand{\isum}[1]{\ensuremath{\sum\limits_{#1=-\infty}^{+\infty}}}

\newcommand{\eref}[1]{(\ref{#1})}

\newcommand{\llrr}[1]{\ensuremath{\left( #1\right)}}
\newcommand{\llrrq}[1]{\ensuremath{\left[ #1\right]}}

\begin{document}

\title{Quantum state transfer via Bloch oscillations}
\author{Dario Tamascelli}
\email{dario.tamascelli@unimi.it}
\thanks{Corresponding author}
\affiliation{Dipartimento di Fisica, Universit\`a degli Studi di Milano, I-20133 Milano, Italy}
\affiliation{Institut f\"ur Theoretische Physik \& IQST, Albert-Einstein-Allee 11, Universit\"at Ulm, Germany}
\author{Stefano Olivares}
\affiliation{Dipartimento di Fisica, Universit\`a degli Studi di Milano, I-20133 Milano, Italy}
\affiliation{CNISM UdR Milano Statale, I-20133 Milano, Italy}
\author{Stefano Rossotti}
\affiliation{Dipartimento di Fisica, Universit\`a degli Studi di Milano, I-20133 Milano, Italy}
\author{Roberto Osellame }
\affiliation{Istituto di Fotonica e Nanotecnologie, Consiglio Nazionale delle Ricerche, Piazza Leonardo da Vinci 32, I-20133 Milano, Italy}
\affiliation{Dipartimento di Fisica, Politecnico di Milano, Piazza Leonardo da Vinci 32, I-20133 Milano, Italy}
\author{Matteo G. A. Paris}
\affiliation{Dipartimento di Fisica, Universit\`a degli Studi di Milano, I-20133 Milano, Italy}
\affiliation{CNISM UdR Milano Statale, I-20133 Milano, Italy}
\date{\today}
\begin{abstract}
The realization of reliable quantum channels, able to  transfer a
quantum state with high \emph{fidelity}, is  a fundamental step in the
construction of scalable quantum devices. In this paper we describe a
transmission scheme based on the genuinely quantum effect known as Bloch
oscillations. The proposed protocol makes it possible to carry a quantum
state over different distances with a  minimal engineering of the
transmission medium and can be implemented and verified on current
quantum technology hardware.  \end{abstract}
\maketitle

\section*{Introduction}
The possibility of transferring, or sharing, a quantum state between
different parties of a quantum network is of fundamental importance in
quantum computation and communications systems. In  solid-state
implementations of quantum devices, for example, several small units
need to be connected in order to share information among them, much the
same way current (classical) computers components are. The realization
of reliable channels, able to the transfer of quantum information with
high \emph{fidelity}, is therefore a fundamental step in the
construction of a scalable quantum computer \cite{divincenzo00}. The
latter, in turn, hold the promise of speeding up the solution of certain
problems perceived as difficult on a classical computer \cite{childs10}
and of enabling controlled simulations of the behavior of complex
quantum systems \cite{feyn82,georgescu14}.  Different quantum state
transfer (QST) schemes have been proposed in the last decade. The range
of systems that can be engineered for the task  is quite
large \cite{bose07,kay10}. However, on
ground of physical implementability and scalability, protocols
that: i) avoid interactions with the system except at
initialization and read-out; ii) are time-independent Hamiltonian,
are to be preferred \cite{kay10}.  A periodical switching of a control field can
be also of interest in view of an almost dispersionless transport over
long distances \cite{childs10,breid07}.
Recently, an experimental verification on
waveguide lattices \cite{chris03} of the perfect-state transfer
protocols proposed in \cite{peres85,christandl04}  and \cite{kay09} has
been reported in \cite{perez13} and \cite{weimann14}.
\par
Here we propose a protocol to exploit Bloch oscillations \cite{bloch29} in order 
to achieve nearly optimal state transfer.  We use the probability of transfer of an
information carrier between two different regions of a transmission line
as a figure of merit and study the trade-off between the amount of
resources used to prepare the initial state of the carrier and the
transfer probability. 
\par
Our protocol requires a minimal engineering of the channel, consisting
in the introduction of an externally tunable temperature
\cite{pertsch99} gradient, or an electric field \cite{peschel98}. It
offers the remarkable possibility of changing the transmission distance
without modifying the geometry of the device. This feature represents a
major innovation compared to  previous QST protocols.
The existing QST schemes are static: a given device, or \emph{channel},
is able to transfer information only between two fixed endpoints. Our
proposal, though based on a time-independent Hamiltonian, opens instead
the possibility of dynamically reconfiguring the ``routing'' of the
transmitted quantum information, a fundamental requirement in any
quantum information processing device. This innovative  scheme  can be
implemented and verified on current photonic lattices technology
\cite{peschel98,pertsch99, sansoni10, corrielli14} and can lead to the
realization of the first reconfigurable QST device for photonic qubits.
Moreover, it can find applications in all-optical switching of light in
communication systems \cite{dawes05}.

\section*{Results}
{\bf The model.} The system we deal with is  a 1D lattice. We indicate each site of 
the lattice by \ket{n}, $n \in \mathbb{Z}$.
The system Hamiltonian is given by (we set $\hbar = 1$):
\begin{equation}	\label{eq:H_TB}
H_0 = -\frac{\Delta}{4} \isum{n} 
\big(  \ketbra{n+1}{n} + \ketbra{n}{n+1} \big),
\end{equation}
where $\Delta$ is the coupling between next neighbor sites. The same
Hamiltonian also governs the dynamics of the Tight Binding Model (TBM)
\cite{harrison89}  and of waveguide-array systems \cite{chris03}.  The
corresponding Schr\"{o}dinger equation is easily solved once we consider
the representation in Bloch waves  \cite{hartmann04,holthaus96}, namely:
\begin{equation} \label{eq:bloch_waves}
\ket{\kappa} = \sqrt{\frac{d}{2\pi}} \isum{n} e^{in \kappa d}\;\ket{n},
\end{equation}
where $d$ is the distance between each site composing the chain, with
the (quasi-)momentum $\kappa$ confined to the Brillouin zone $-\pi/d \le
\kappa \le +\pi/d$. These states are eigenstates for the Hamiltonian
\eqref{eq:H_TB} with  eigenvalues \begin{align} \label{eq:dispersion}
E(\kappa) = -\frac{\Delta}{2} \cos(\kappa d).
\end{align}
Equation (\ref{eq:dispersion}) expresses the well-known momentum/energy dispersion relation in lattices.

\begin{figure}[h!]
\includegraphics[width=0.39 \columnwidth]{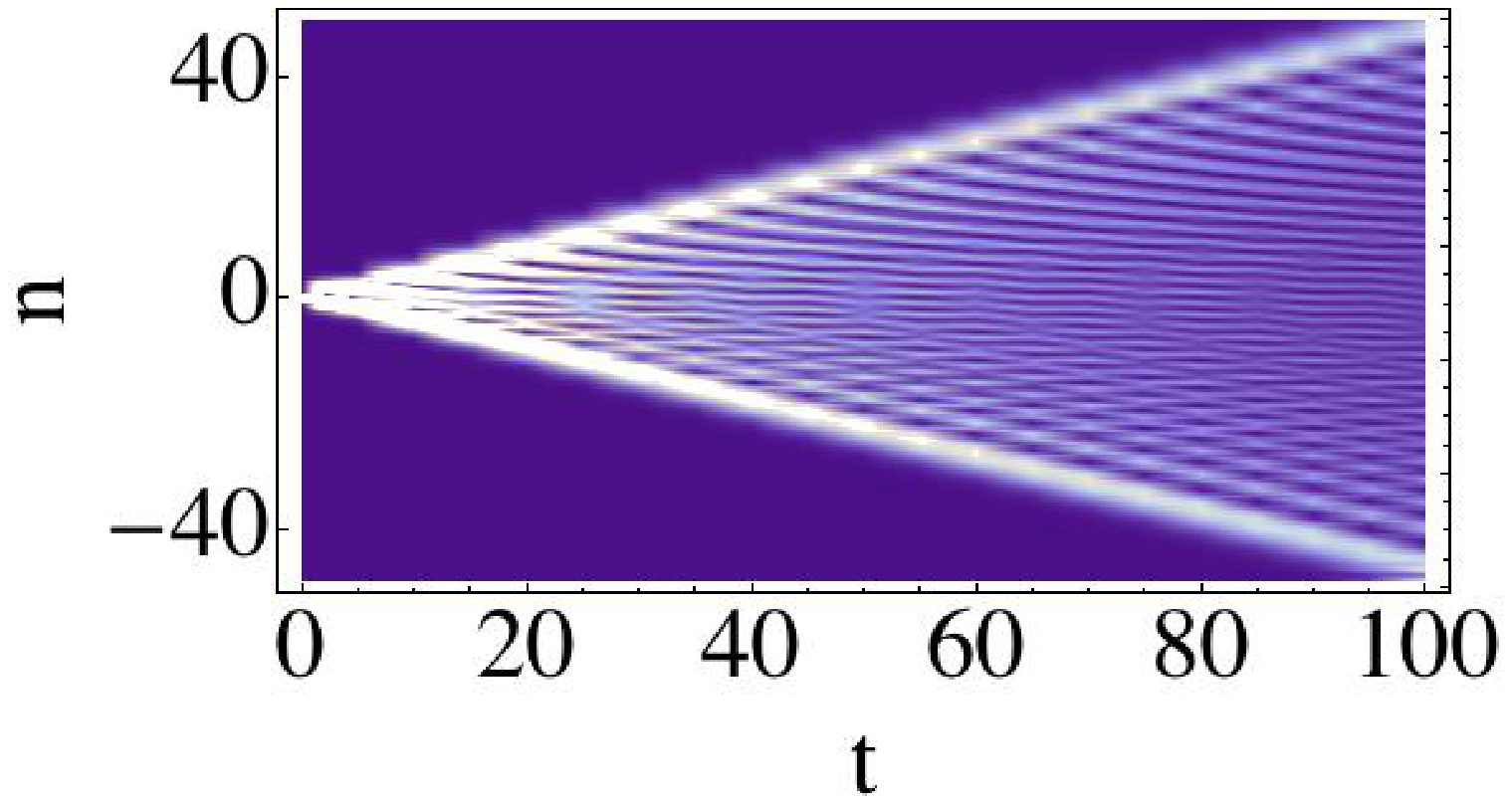}
\includegraphics[width=0.39 \columnwidth]{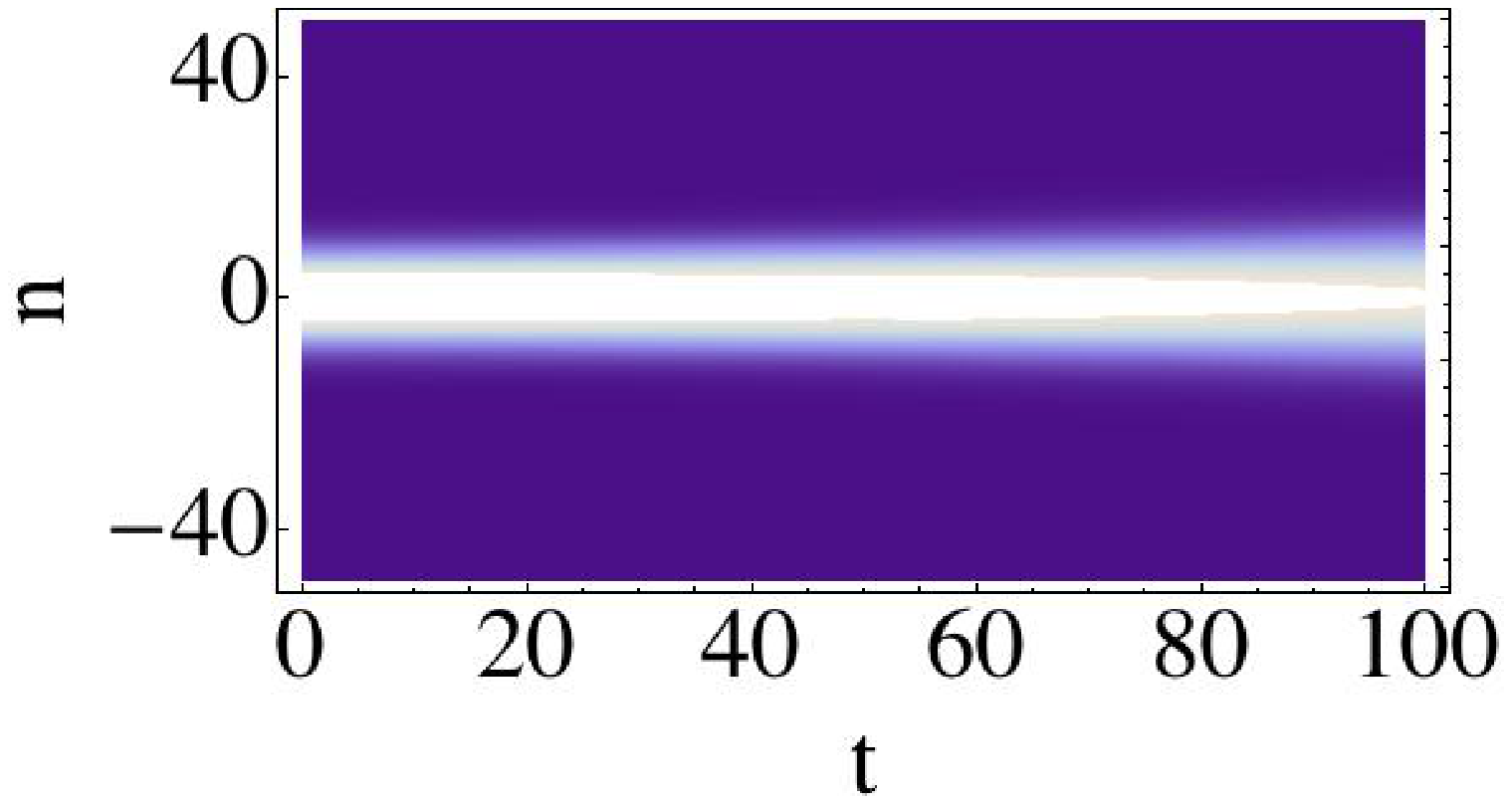}
\includegraphics[width=0.39 \columnwidth]{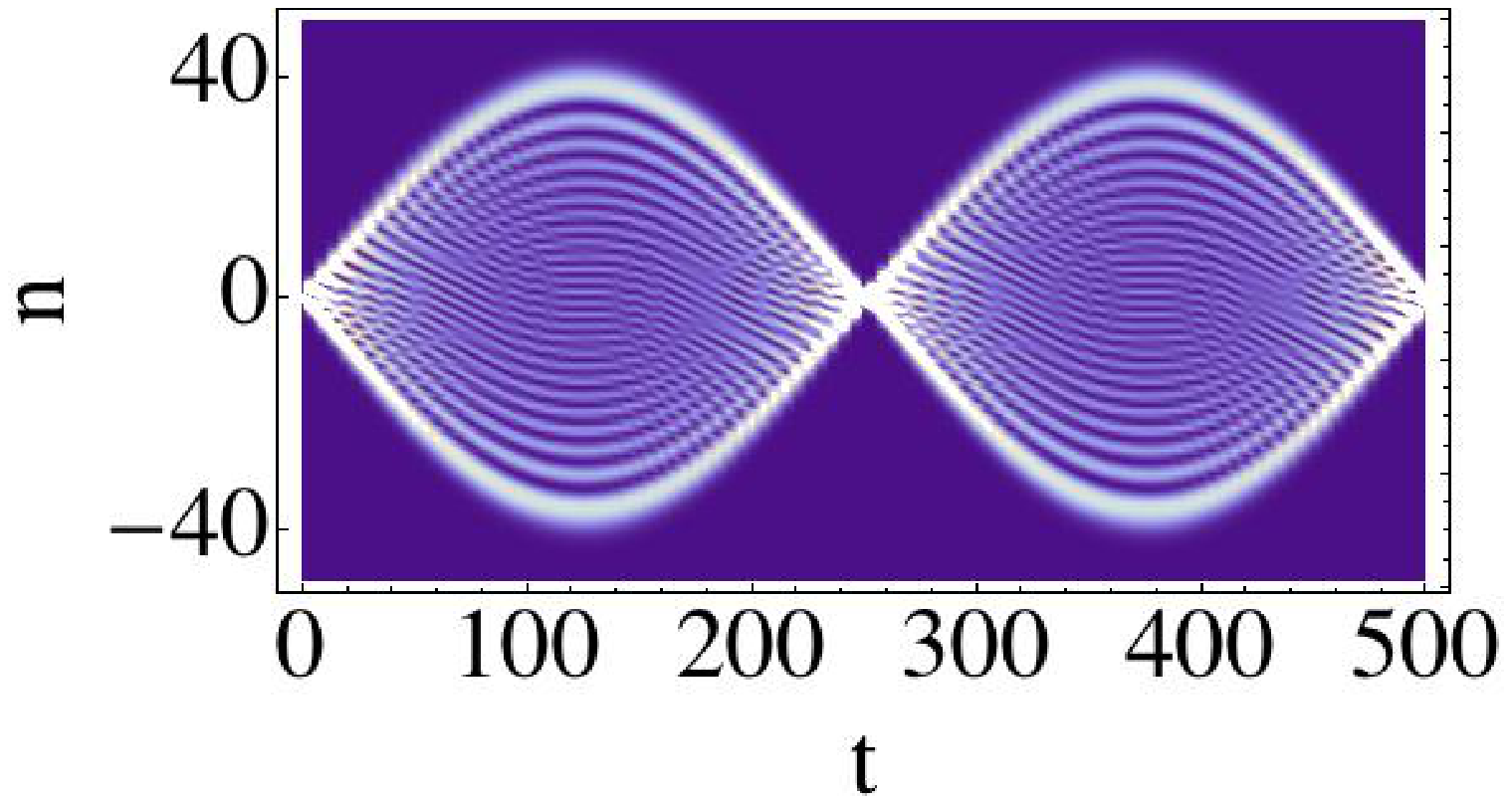}
\includegraphics[width=0.39 \columnwidth]{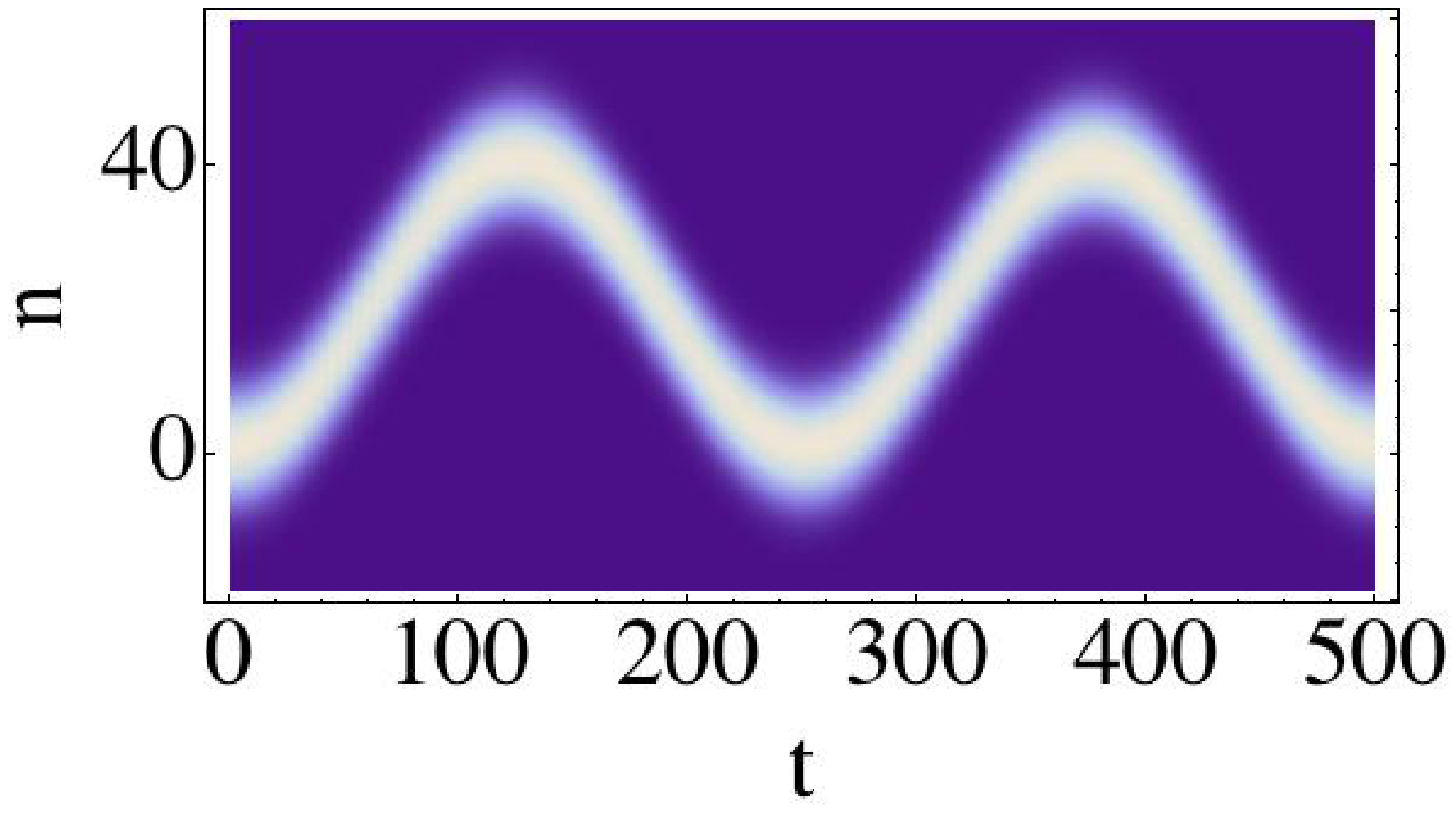}

\caption{
Evolution of the probability distribution
$\left |\braket{n}{\psi(t)} \right |^2$ (density plot) 
for $\Delta  =1, \ F  =-1/40,\ 
\beta = 0.01 $ and for different initial conditions and system 
Hamiltonians. (Top left): Sharp initial condition $\ket{\xi_0^\delta}$ 
\eref{eq:sharp_initial_cond}, $H = H_0$ \eref{eq:H_TB}; 
(top right): Gaussian distribution \ket{\xi_0^G} \eref{eq:gauss_initial_cond},  
$H=H_0$; (bottom left): Sharp condition $\ket{\xi_0^\delta}$, $H = H_B$ 
\eref{eq:H_Bloch}; (bottom right): 
Gaussian wavepacket \ket{\xi_0^G}, $H = H_B$. \label{fig:free} }
\end{figure}

The evolution of a generic state $\ket{\xi_0} = \sum_n c_n(0)\ket{n}$ 
is obtained by  the propagator $U(t) = \exp\llrr{-i H_0 t}$. In this simple setting we have $\bra{n} U(t) \ket{n'} = i^{n-n'} J_{n'-n}(\frac{t \Delta}{2})$,  where $J_n(x)$ is the $n$-th Bessel-J function of the first kind. 
In Figs. \ref{fig:free} and we show the evolution of two different
initial conditions: a (sharp) localized condition and a Gaussian
wavepacket, both centered in the site labelled by $0$, namely:
\begin{align} 
	&\ket{\xi_0^\delta}=\ket{0} \qquad \text{or} \label{eq:sharp_initial_cond}\\
	&\ket{\xi_0^G}=\left( \frac{2\beta}{\pi}\right)^{1/4} \isum{n} e^{-\beta n^2} \ket{n} .\label{eq:gauss_initial_cond}
	\end{align}
If our task is to transfer the excitation/electron/photon from the initial position $n_I=0$ to a final position, or \emph{target site},  $p$ ($p=40$ in the examples shown in the figures), this simple setup is completely uneffective. When starting from the  sharp initial condition \eref{eq:sharp_initial_cond} the probability of reaching the target site decreases polynomially with the distance $|p|$.  Starting from \eref{eq:gauss_initial_cond} leads to a diffusive behavior, since the initial momentum has mean $0$ and variance $2\beta$ \cite{hartmann04}. 
%
\\[5pt]{\bf Bloch oscillations.} We now add a linear potential to the Hamiltonian \eref{eq:H_TB}, mimicking the action of a ``force'' trying to pull the excitation in the desired direction. This could induce inter-band transitions \cite{gluck02,mandelik03}; however, since we are going to consider initial states with negligible transverse moment and small values of the force parameter,  transitions to higher bands can be safely neglected \cite{longhi06}. We can therefore introduce the following single-band Hamiltonian:
\begin{align} \label{eq:H_Bloch}
H_B = &
H_0+ F d \isum{n} n \ketbra{n}{n}.
\end{align}
The eigenvalues of $H_B$ are $E_B^m = m d F$ and the corresponding eigenstates are the Wannier-Stark states $\ket{\Psi_B^m},\ m = 0, \pm 1,\ldots$, where
\begin{align}\label{eq:WS}
\Psi_B^m(\kappa) &= \braket{\kappa}{\Psi_B^m} = \sqrt{\frac{d}{2\pi}} e^{-i \llrrq{m \kappa d + \gamma \sin(\kappa d)}}, 
\end{align}
and $\gamma = \Delta/(2d F)$ \cite{hartmann04}. The propagator, in the Bloch basis \eref{eq:bloch_waves}, is
\begin{align}
\braoket{\kappa'}{U(t)}{\kappa} = e^{-i \gamma \llrrq{\sin(\kappa'd) - \sin(\kappa d)}} \delta(\kappa'-\kappa+ F t).
\end{align}
The quasi-momentum $\kappa$ is changed by the \emph{force} as $\kappa(t) = \kappa(0) - F t$ . On the other side, the group velocity $v_g(\kappa)$ of the wave, is defined through the dispersion relation \eref{eq:dispersion} by
\begin{align} \label{eq:group_velocity}
v_g(\kappa) = -\frac{d}{d\kappa} E(\kappa) = \frac{d \Delta}{2} \sin(\kappa d);
\end{align} 
it changes sign every time the quasi-momentum $\kappa$ reaches the
boundaries of the first Brillouin zone, leading to Bloch oscillations
\cite{bloch29}. The evolution of the initial states
\eref{eq:sharp_initial_cond} and \eref{eq:gauss_initial_cond} in the
presence of a potential is illustrated in Fig.~\ref{fig:free}.  The
appearance of the \emph{breathing modes} (see lower left panel) is a
consequence of the flat momentum spectrum corresponding to the sharp
initial condition \eref{eq:sharp_initial_cond}. The components having
absolute initial momentum $|\kappa| \approx 0(\pm \pi)$ travel the
furthest along the chain, reaching a distance $\Delta/F$. Their speed,
initially close to  $d \Delta/2 \sin(0(\pm \pi)) = 0$, increases in
module until it reaches the maximum value $d \Delta/2$ at $t = \pi/2 d
F$. Then the velocity decreases in modulus until $t=\pi/dF$ when it
changes sign (Bragg reflection). The initially fastest components of the
wave packet, corresponding to  $|\kappa| \approx \pi/2$, get Bragg
reflected sooner, and are confined in a region $(-\Delta/2F,\Delta/2F)$.
The presence evenly distributed initial positive and negative momenta,
leads moreover to an even spreading of the wavepacket over both the
positive and negative axes; this leads to a further halving of the
probability of reaching a target site $p$. 
\par
\begin{figure}[h!] 
\includegraphics[width=0.44\textwidth]{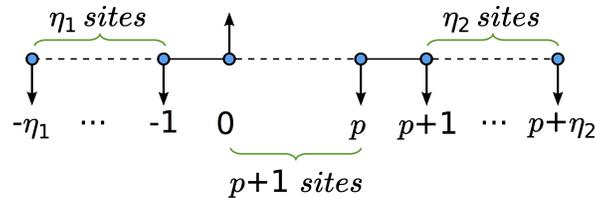}
\caption{\label{fig:stretch}Finite chain representation.}
\end{figure}

On the other side, if we take a Gaussian initial condition of the form
\eqref{eq:gauss_initial_cond}, with  $\beta \ll 1$, the distribution of
the momentum can be  peaked around $\kappa=0$. The wavepacket will now
travel in a definite direction, set by the sign of $F$ and the shape of
the starting Gaussian is preserved during the evolution. In fact,
besides a phase factor $e^{-in\omega_Bt - i\Phi(t)}$ \cite{hartmann04}
the coefficients distribution is the same as in
\eqref{eq:gauss_initial_cond} with the substitution: $n \rightarrow
n-\bar{n}(t)$,  where $\bar{n}(t)$ is the mean of the position
observable $\widehat{n} = \isum {n} n \ketbra{n}{n}$. The second
relevant point is that the center $\bar{n}(t)$ of this Gaussian shape
performs an oscillation with period $T_B = 2\pi/(Fd)$ within the
coordinate space. So, as long as the initial shape is weakly localized
in the coordinate space, we expect that the whole Gaussian shape -
representing the probability distribution for each site - performs an
oscillation with amplitude $ 2 \left|\gamma \right|$.  We point out,
moreover,  that after a half period the coefficients $c_n(t)$  take the
form:
	\begin{align} \label{eq:alternate}
	c_n(t=T_B/2)&=
	 (-1)^n \, e^{-\beta (n + 2\gamma)^2},
	\end{align}
	i.e.  the wavefunction is the same as the initial one but
	shifted of $-2\gamma$ and with alternate phase factors.
	Then, in this toy model, we are able to transfer the excitation
	from a site to another arbitrarily just varying the force acting
	on the system.
\\[5pt]{\bf Constraining the resources.} We now want to understand under which conditions the dynamics on a finite chain approximates properly the one discussed so far. This issue is quite relevant since in any realistic setting the number of lattice sites or waveguides would be limited. Let $p$ once more indicate the target site and suppose that the excitation is initially localized in a neighborhood of the site labelled by $0$. We suppose to attach $\eta_1$ sites before the site labelled by 0 and $\eta_2$ after the one labelled by $p$, as shown in figure \ref{fig:stretch}.
\par
\begin{figure}[h!]
\includegraphics[width=0.5\columnwidth]{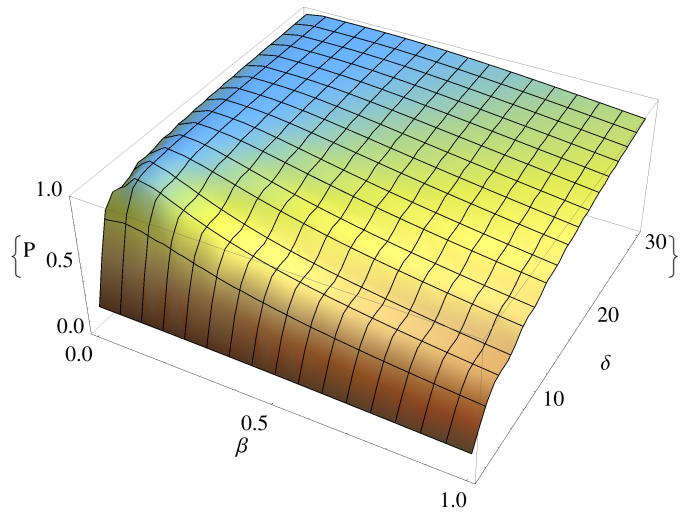}
\caption{\label{fig:beta_delta_QP}
The success probability  ${\cal P}(p\equiv 40,t\equiv T_B/2,\delta)$ as
a function of $\beta$ and $\delta$ for $\Delta/F=-60$.}	
\end{figure}

The total number of sites composing the chain is therefore $c = (p+1) + \eta_1+ \eta_2$.
In order so simplify the notation we introduce the quantities: $l\equiv
-\eta_1$ and $r \equiv p+\eta_2$. Dealing with the finite case the
Hamiltonian governing the system will be: \begin{align}
\label{eq:H_finite}
	H_f  = & -\frac{\Delta}{4} \sum\limits_{n=l}^{r-1} \big(  \ketbra{n+1}{n} + \ketbra{n}{n+1} \big) +Fd\sum\limits_{n=l}^{r} n\ketbra{n}{n}. 
	\end{align}
Of course, the Gaussian superposition
in \eqref{eq:gauss_initial_cond} cannot be extended to an infinite number
of sites, but at  most to the ones composing the chain. Moreover the
state \ket{p} must not appear in the Gaussian superposition: if it were
the case, we would have a non-vanishing initial probability of finding
the spin-excitation in the target site. It is clear then the necessity
of taking the Gaussian superposition truncated in certain interval on
the chain, so that it involves only a restricted number of sites. In
particular we considered a Gaussian superposition symmetrically
truncated with respect to its center. Chosen a truncation parameter
$\delta < \eta_1$ we set the initial state to be: 
\begin{equation}
\label{eq:truncated_sperposition}
	\ket{\xi_G^\delta} = A\sum_{n=-\delta}^{+\delta}  e^{-\beta n^2}\;\ket{n}	,
	\end{equation}
where $A$ is a normalization factor, and investigate how the truncation
affects the transport property. As a figure of merit we employ the
probability of finding the excitation in a neighborhood of the final
site $p$. Upon denoting by 
\begin{equation}
P(n,t) = \left| \braket{n}{\xi_G^\delta(t)}\right|^2 \label{eq:defP}
\end{equation}
the probability of finding the spin excitation in the $n$-th site, we
define the \emph{success probability} as: \begin{equation}
\label{eq:success_probability}
	{\cal P}(p,t,\delta) = \sum\limits_{n=p-\delta}^{p+\delta} P(n,t).
	\end{equation}
Our main aim is to maximize such probability. The  parameters we can
control are the force intensity $F$, the truncation parameter $\delta$
and the Gaussian superposition width $\beta$. The value $F$ is
automatically set once we decide how far the excitation has to travel.
Notice that, if $\delta$ is taken too close to 1 the Gaussian
superposition becomes very similar to the sharp condition, leading to
the unproductive breathing modes shown in the lower left panel of
Fig.~\ref{fig:free}. As already discussed above, moreover, we need $\beta
\ll 1$ in order to have a momentum distribution peaked at $0$.
In our numerical simulations we set  $\eta_1 =  \eta_2= 2 \delta$ to
avoid dangerous edge effects on the evolution of the truncated Gaussian
superposition. As an example we plotted the values of  \emph{success
probability} for $\Delta/F=-60$ in Fig.~\ref{fig:beta_delta_QP} at the
optimal time $t=T_B/2$ (which does not depend on $\beta$ or $\delta$). 
\par
We notice that the mutual dependence of $\beta$ and $\delta$ is clearly
visible in all the region plotted.	 The behavior of the success
probability at the vary of $\delta$ is clearly expected. In fact, as
$\delta$ increases the region on which we collect the squared amplitudes
$c_n(T_B/2)$ covers an always wider part of the whole lattice. Obviously
in the extreme case in which the region covered is the whole lattice the
success probability is exactly $1$. It is clear that exists a value for
$\beta$ for which the success probability is almost independent on
$\delta$ and very close to 1.  The value we find is about $\beta =
0.01$. For  $\delta=5$  the success probability is already around 0.9
and reaches the value 1 for $\delta=16$.  As we deal with a finite
chain, it makes sense to take the lowest value of $\delta$ that makes
the success probability larger than an assigned threshold value.

\section*{Discussion}
\begin{figure}[h]
\subfigure[]{\label{fig:figx} \includegraphics[width=0.45 \columnwidth]{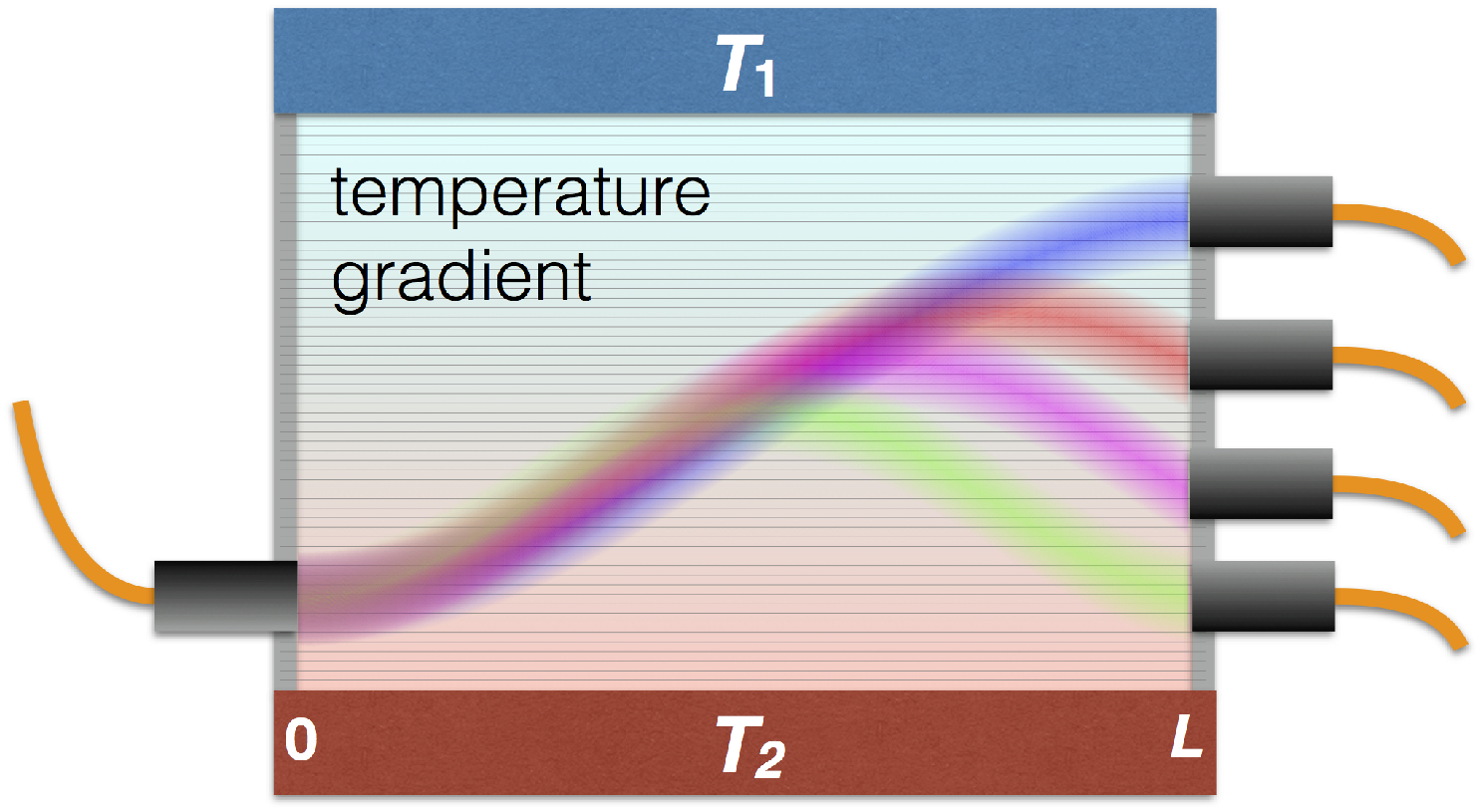}}
\subfigure[]{\label{fig:figy} \includegraphics[width=0.45 \columnwidth]{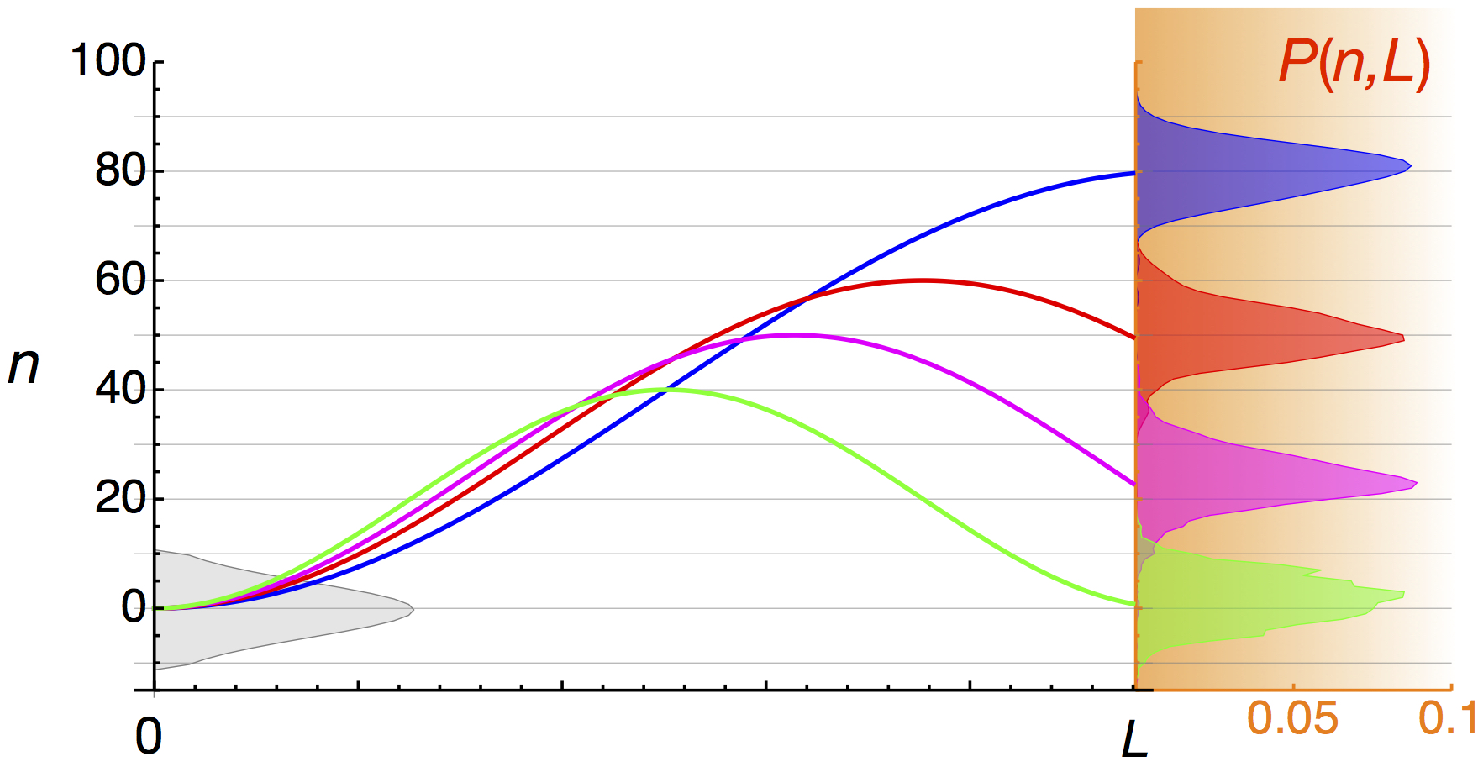}}
\caption{\label{fig:transSetup} (a) A possible scheme of the experimental implementation with a simulation of the propagating signals from left to right. We set the coupling constant $J=1$ and the initial condition is a truncated Gaussian with $\beta = 0.01, \delta = 10$ centered at $n=0$. The colors correspond to different values of $F$, determined by the temperature gradient between $T_1$ and $T_2$: $F= 1/80$ (blue), $F=1/60$ (red), $F=1/50$ (magenta), and  $F=1/40$ (green). (b)~The left plot shows the mean position of the wavepackets simulated in the scheme (a) as a function of the distance from the input point; $n$ refers to the number of the waveguide. On the right we show the intensity profiles $P(n,L)$ of the wavepackets at the output of the waveguide array, as obtained through numerical simulations. The grey profile in the left plot refers to the  the intensity profile of input truncated Gaussian where we used the same scale as for $P(n,L)$ (not shown for the sake of clarity).
}
\end{figure}

Now that we know in good approximation the dynamics of an excitation on
a finite chain we can use the results obtained to perform an information
transfer. The  Hamiltonian \eref{eq:H_TB}, with $n=1,2,,\ldots,N$ is
equivalent to the one governing the propagation of the light in an array
of $N$ evanescently coupled optical waveguides \cite{chris03}. The
introduction of a linear potential will lead to solutions of the form
\eref{eq:WS}, i.e. the Wannier-Stark states \cite{peschel98,hartmann04}.
The external force acting on the lattice would be implemented by a
linear gradient in the effective refractive index of the waveguides.
This could be realized statically, but also dynamically, e.g. imposing a
temperature gradient in the substrate and exploiting the thermo-optic
effect \cite{pertsch99}. In Fig.~\ref{fig:figx} we show the scheme of a possible
implementation and the simulation of the propagating signals. We
plot in Fig.~\ref{fig:figy} the mean position of the wavepackets simulated
in (a) during the propagation through the waveguide array together with the intensity profiles $P(n,L)$, defined as in \eref{eq:defP},  at the output:
it is evident that the output states are well distinguishable and only slightly deformed with
respect to the input [the grey profile in Fig.~\ref{fig:figy}]. In both figures the time parameter $t$ has been replaced by the length $L$ of the waveguide array.
\\In this setting, we have also an additional degree
of freedom, namely the polarization of the light propagating in the
array.  This additional two-level degree of freedom does not interfere
with the motion of the light in the lattice, as long as the coupling
between the guides is polarization independent. We can then suppose that
the light propagating through the lattice preserves its polarization.
Light is thus carrying a qubit of information, encoded in its
polarization state, from an initial region of the lattice to the target
one. By modulating the external force the qubit could be displaced and
dynamically redirected in different regions of the lattice.
\par
The polarization state is analogous to a spin-1/2 state, characterized
by the eigenstates $\ket{\downarrow_p}$ and $\ket{\uparrow_p}$, with
respect to the Pauli operator $\sigma_z^e$. The necessary Hilbert space
for such a degree of freedom is $\mathcal{H}^P = \mathrm{span} \lbrace
\ket{\downarrow_p}, \ket{\uparrow_p} \rbrace $. The overall Hilbert
space needed for the complete description of the system is:
$\mathcal{H}^e \equiv \mathcal{H}^C \otimes \mathcal{H}^P$. 
As the initial wavefuction we use $\ket{\xi_G^\delta}$ defined above 
with the addition of the polarization
state, i.e. 
\[
\ket{\nu} \equiv \ket{\xi_G^\delta} \otimes \ket{s_p}
\]
where $\ket{s_p} \in \mathcal{H}^P$ is the polarization state. Since in
non-birefringent media the polarization of the light is not affected
during the propagation, the Hamiltonian governing the motion of the
electron through the lattice is $H_e =  H_f  \otimes \mathbb{I}_P$,
where $\mathbb{I}_P$ is the identity operator on the Hilbert space
$\mathcal{H}_P$. Under the influence of this Hamiltonian the light
particle can carry quantum information under the form of its
polarization from a site to another following the dynamics discussed
above.

Waveguide arrays supporting such a dynamics could be fabricated by
femtosecond laser writing \cite{dellaValle09}. This technique allows to
directly inscribe high quality waveguides in glass substrates,
exploiting the non-linear absorption of ultrashort laser pulses. This
technology has widely proved its capabilities in producing complex
three-dimensional waveguide arrays, able to reliably implement or
simulate diverse quantum dynamics \cite{longhi09,szameit10} and, in
particular, the Bloch oscillations of light \cite{chiodo06, dreisow11,
corrielli13}. Furthermore, it has been recently shown that femtosecond
laser written circuits are specially suitable for the manipulation of
polarization-encoded qubits, thanks to the relatively low birefringence
that characterizes the waveguides fabricated with this technique
\cite{sansoni10} and the possibility of fabricating polarization
insensitive devices \cite{sansoni12}.

A truncated Gaussian input state can be experimentally implemented in
free space by using hard apertures together with a cylindrical telescope
and a microscope objective to launch light in the array
\cite{pertsch02}. Although this method may be extremely effective to
characterize the device and to demonstrate the quantum transfer effect,
it may not be the best choice when this device will be used in an
integrated environment, e.g. inside a quantum computer. In that case, it
would be more appropriate to exploit engineered photonic lattices to
transform the single mode of an incoming waveguide into a truncated
Gaussian state that will constitute the input state of the quantum
transfer device. In particular, the engineered photonic lattice could
consist in a linear array of waveguides with tailored coupling
coefficient, e.g. mirroring a $J_x$ matrix \cite{weimann15}. A single
waveguide excitation will produce a Gaussian output distribution. By
connecting only a truncated set of such waveguides to the quantum
transfer chip one would achieve the desired input state. The reversed
solution can be implemented to collect the target state.
\\[5pt]\indent While Bloch oscillations are well known since the early stages of
quantum mechanics,  here we propose a way to take advantage of them for
the task of quantum state transmission. We showed that the constraints
imposed by the finiteness of the resources available for the preparation
of the initial state induce a minor lowering of the protocol efficiency.
The minimal amount of engineering required to implement the system
Hamiltonian, and to prepare the initial state, make the realization of
the protocol feasible with current quantum technology. Our results pave
the way for further investigations concerning the system, such as  the
effects of noise \cite{benedetti13,rossi14} 
and imperfections \cite{dt13} on the transmission probability. 

\section*{Acknowledgments}
D.T., S.O., S. R. and M.G.A.P. acknowledge financial support by EU through the Collaborative Project QuProCS (Grant Agreement 641277) and by UniMI through the H2020 Transition Grant 14-6-3008000-625 and the grant ÒSviluppo UniMi 2015Ó. R.O. acknowledges financial support by EU through the Collaborative Project QWAD (Grant Agreement 600838).

\section*{Author Contributions}
D.T. and S.O. conceived the protocol. D.T., S.R. and S.O. performed the numerical and analytical analysis.
D.T., S.O., R.O. and M.G.A.P. studied the feasibility of the protocol with current waveguide technology.
All authors contributed to the preparation of the manuscript.


\end{document}